# Direct correlation of intravoxel incoherent motion imaging based on implicit neural representations and histopathology in pediatric Wilms' tumors: can it differentiate between subtypes?

G.C. Arends* [1], M. Fitski* [2], J. Harbers [2], J.N. van der Beek [2,3], R.R. de Krijger [2,4], M.M. van den Heuvel-Eibrink [2], A.F.W. van der Steeg [2], F.F.J. Simonis [5], E. Versteeg [1], A.S. Littooij [2,3], C.M.W. Tax [1]

** These authors contributed equally to this study*

[1] Center for Image Sciences, University Medical Center Utrecht, Utrecht, The Netherlands

[2] Princess Máxima Center for Pediatric Oncology, Utrecht, The Netherlands

[3] Department of Radiology and Nuclear Medicine, University Medical Center Utrecht/Wilhelmina Children's Hospital, Utrecht University, Utrecht, The Netherlands

[4] Department of Pathology, University Medical Center Utrecht, Utrecht, The Netherlands

[5] Multi-modality Medical Imaging, University of Twente, Enschede, The Netherlands

**Background:** Wilms tumor (WT) is a heterogeneous pediatric renal malignancy. Invasive procedures to establish histology prior to pre-operative treatment are not recommended by the International Society of Pediatric (SIOP) Oncology Renal Tumor Study Group, yet different tissue components are associated with different responses to therapy (SIOP Umbrella protocol). Intravoxel incoherent motion (IVIM) magnetic resonance imaging (MRI) may improve subtype differentiation before surgery as a noninvasive radiographic biomarker by capturing both diffusion and perfusion characteristics, as opposed to the conventionally-used apparent diffusion coefficient (ADC). The objective of this study is to evaluate the discriminative performance of IVIM MRI in differentiating WT subtypes. IVIM provides quantitative parameters that disentangle the diffusion coefficient (D), incoherent blood flow (pseudodiffusion, D*) and perfusion fraction (f).
**Materials and methods:** This retrospective study included thirthy children with WT. Standardized 1.5T MRI with 10-b-value diffusion weighted imaging (DWI) was performed. IVIM parameters were estimated using a self-supervised implicit neural representations network and were correlated with histopathology via a patient-specific 3D-printed cutting guide. Statistical analyses included mixed models, receiver operating curve analysis, and logistic regression.
**Results:** Thirty lesions were MRI–histopathology correlated on slice-level. Significant subtype differences were observed for D* stromal versus epithelial ($p = 0.024$), regressive tumors ($p = 0.031$) and mixed WT tumors ($p = 0.043$). ADC showed no significant subtype differences.
**Conclusions:** Direct MRI–histopathology correlation enabled differentiation between stromal subtype and epithelial, mixed and regressive subtype on macroscopic slice level using D*. Future research should focus on data addition and optimizing equal representation of different tissue types.

## Introduction

Wilms' tumor (WT) is the most common pediatric renal malignancy, comprising roughly 85–90% of all childhood renal tumors [1], [2], [3] . It typically arises in children between 1 to 4 years old [2], [4], [5]. Known for its heterogeneous tissue characteristics, WT consists of varying proportions of blastemal, stromal, and epithelial elements, which react differently to therapy. The current SIOP-RTSG 2016 UMBRELLA protocol recommends pre-operative chemotherapy without prior histological confirmation through a biopsy, due to risks of tumor rupture and subsequent need for radiotherapy, and sampling errors [6], [7] . Consequently, histopathological evaluation and risk stratification are performed after surgery, which determines adjuvant therapy[8-10] . This general neoadjuvant therapy approach may lead to over- or undertreatment. Thus, there is a need for reliable methods of preoperative subtype identification of the whole tumor to support a more personalized neoadjuvant therapy strategy and to improve clinical outcomes.

To facilitate this preoperative subtype identification, diffusion weighted imaging (DWI) has shown to be promising in providing non-invasive radiographic biomarkers, particularly using the apparent diffusion coefficient (ADC) [11], [12], [13], [14], [15] . Previous studies have shown that stromal-type WT show higher ADC values compared to other histopathological subtypes, correlating with its lower cellular density [11], [16], [17] . However, ADC on its own lacks microstructural specificity and the ability to differentiate between more cellular subtypes such as blastemal and epithelial tissue. A blastemal component typically has a dense cellular structure, whereas stromal cell areas are less dense and more perfused [11], [17] . To overcome this limitation, intravoxel incoherent motion (IVIM) could offer a more specific approach by separating the combined effect of diffusion and perfusion [18], [19], [20] , providing insight in both cellular microstructure and microvascular perfusion [21], [22], [23] . IVIM provides quantitative parameters that separate the diffusion coefficient (D), incoherent flow in capillaries (pseudodiffusion, D*) and perfusion fraction (f). These parameters have the potential to capture distinct biological properties in data analysis, possibly allowing improved differentiation of WT subtypes. This suggests that IVIM MRI could enhance preoperative differentiation between WT subtypes based on their microstructural and perfusion profiles.

Despite this clinical hypothesis, accurate IVIM parameter estimation is challenging in pediatric renal tumors due to low signal-to-noise ratio (SNR), and limited scan time. Especially f and D* have shown unstable fitting behavior as these parameters are particularly sensitive to the exponentially decaying signal component due to diffusion weighting [24]. Therefore, this work uses implicit neural representations (INRs) for IVIM parameter estimation, exploiting their ability to improve DWI parameter estimation under low-SNR conditions [25]. Our earlier results in children [26] and recent work in adults has demonstrated that this approach is also applicable to IVIM fitting in tumors [27] . INRs leverage spatial correlations to produce a continuous representation, in contrast to voxel-wise fitting methods such as linear least-squares [28] or deep learning approaches [29], [30] . Furthermore, this self-supervised neural network approach eliminates the need for training data and retraining for specific DWI acquisition protocols.

To study the value of IVIM parameters as an imaging biomarker, tumor components on MRI need to be correlated reliably with corresponding microscopic histopathology. To address this, we use patient-specific 3D-printed tumor cutting guides to reliably align imaging and histopathological specimens. This approach improves spatial accuracy of all correlated data, ensuring that IVIM parameters are correlated to corresponding tumor regions [16].

The aim of this retrospective study on prospectively collected data was to evaluate the discriminative performance of IVIM parameters in differentiating WT types, using 3D-printed cutting guides and state-of-the-art IVIM fitting methods to directly correlate histology slices to the imaging slices.

## Materials and methods

**Patients**

The Institutional Review Board of the University Medical Center Utrecht approved this retrospective study [16]. The need for informed consent was waived because the study used previously collected research, clinical and imaging data obtained as part of the UMBRELLA study and posed no additional risk. A dataset of 56 pediatric patients with a renal tumor was used for this study (Figure 1). Data were prospectively collected in the Princess Máxima Center between May 2021 and May 2023. Patients were eligible for inclusion if they had undergone a preoperative MRI-scan, a cutting guide was made and if the correlation was successful. The pre-operative MRI-scan had to be performed within 10 days before total nephrectomy, without any treatment administered between imaging and surgery. Patients were excluded if a partial nephrectomy was performed, if the pathological diagnosis was not a WT, if less than 10 b-values were scanned, and if there was a difference in number of slices between the T1 scan and the T1 scan after intravenous gadolinium contrast agent.

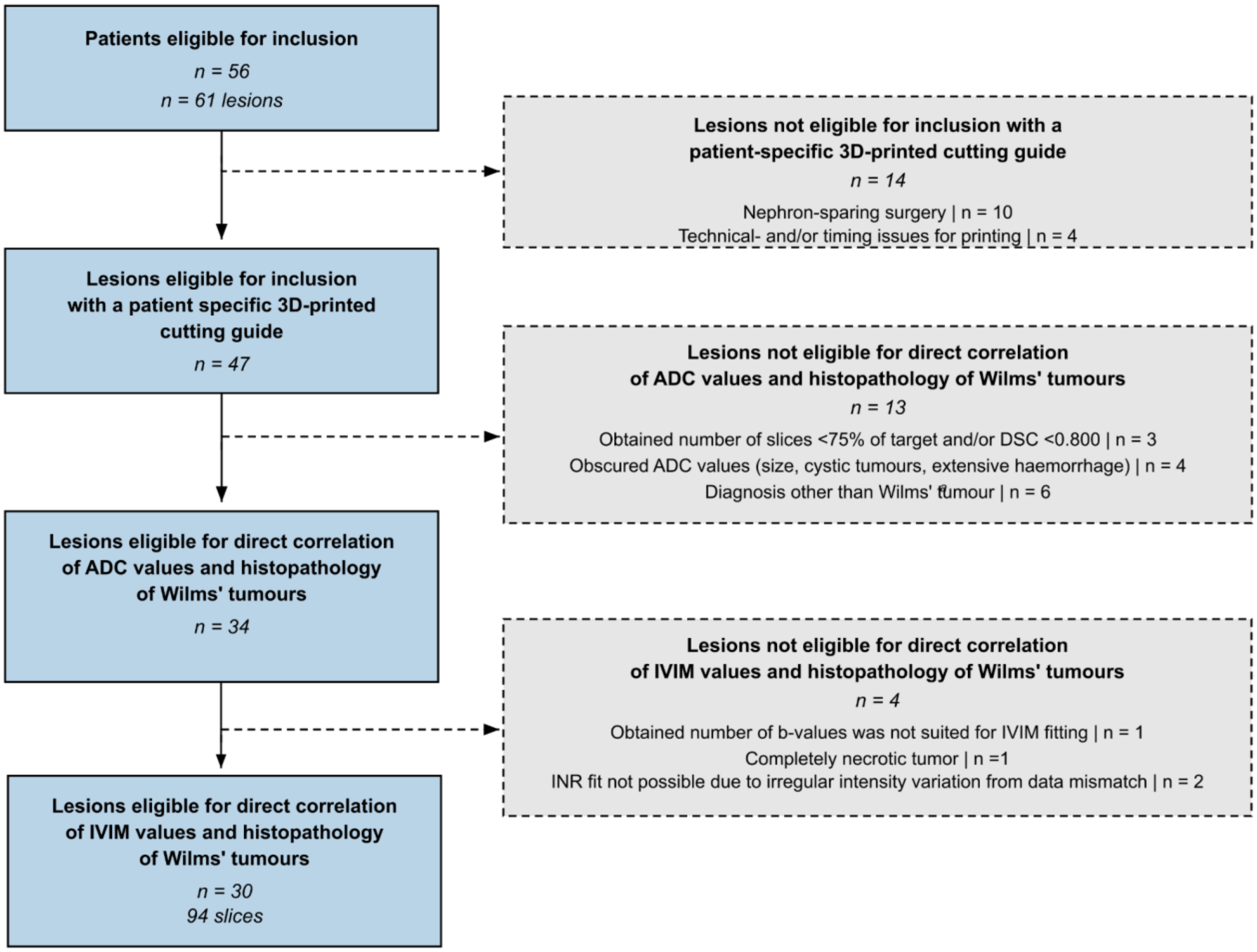

*Figure 1: Flow diagram of the lesions included and excluded in this study.*

## MRI acquisition and analysis

All MRI scans were performed using a 1.5-T scanner (Ingenia, Philips Medical Systems). A standard scan protocol was used including T1W (2-D ultrafast spoiled GRE with fat suppression) with and without intravenous gadolinium contrast, T2W (MultiVane/TSE with fat suppression) and DWI (2-D single-shot spin echo with spectral fat saturation) sequences. The DWI acquisition was performed with 10 b-values (0, 25, 50, 100, 150, 200, 250, 500, 800, 1000 s/mm$^2$). According to standard-of-care practice, children were examined while awake, sedated, or under general anesthesia, based on their level of cooperation. Hyoscine butylbromide (Buscopan; Sanofi) at a dose of 0.4 mg/kg body weight to reduce peristaltic artifacts (maximum of 5 or 10 mg, depending on age) and Gadobutrol (Gadovist; Bayer B.V.) at a dose of 0.1 mL/kg body weight were administered.

## Histopathological review

We optimized and protocolized the use of 3D-printed cutting guides in earlier work [16]. The design of the cutting guide enables the pathologist to maintain consistent slice thickness (5 mm) and to reproduce the orientation of the MRI-DWI for the resulting macroscopic slices. The relative amounts of stromal, epithelial, and blastemal components were documented at the level of each included slice and included an overall annotation of the component type of the slice, following UMBRELLA definitions.

**DWI processing for IVIM**

The INR framework was adapted to IVIM fitting and was trained on an NVIDIA Titan X GPU for 300 epochs using a Mean-Squared-Error loss with a learning rate scheduler (lr = 0.001, factor = 0.99). The specific INR network consisted of a multi-layer perceptron with three hidden layers with size 2048. The total amount of positional encodings was set to 2500. Simulated data was used to validate the performance of INR fitting accuracy with respect to other fitting methods using the Pearson correlation coefficient (r) and root-mean-square error (RMSE) (see supplementary information for methods or Arends et al [26] ). The standard deviation of the INR was optimized via evaluation of fitting performance on simulated data, which was then fixed at 8 for all subsequent INR fittings. A tumor segmentation based on the T2 scan was used during fitting to only include tumor tissue in the fitting process. The code is publicly available through github at: https://github.com/garends3/IVIM_INR, which also provides information on used python packages. ADC values were calculated from the DWI acquisitions with b-values 0 and 1000 s/mm$^2$.

Given the potential distorted measurements of IVIM-values in cystic, hemorrhagic and/or necrotic components, a subtraction of pre- and post-gadolinium T1 scans was performed per patient to mask viable tumor regions based on the method of Littooij et al [13] . The pixels of the subtracted dataset with an enhancement similar to or higher than the spinal erector muscles were used as a threshold. The mean pixel value of the muscles was extracted from the T1 subtraction scans using a manually selected ROI in Python (Version 3.10.16). Together with the T2 segmentation, this subtracted T1 data was used to create a mask of the tumor without necrotic, hemorrhage or cystic elements. This mask was used for statistical analysis.

**Statistical analysis**

For every MRI slice corresponding to a pathology slice, the difference of IVIM-derived parameter values (median and mode) between stromal-, epithelial-, blastemal-, mixed- and regressive-components was analyzed using a mixed model analysis. Results were considered significant if the p-value was below 0.05. Diagnostic performance was further evaluated using logistic regression models on specific WT subtypes. Similar analysis was performed on median ADC values. Statistical analyses were performed in Python (version 3.11.9) using scikit-learn (version 1.9.0).

## Results

**Patient characteristics**

A total of 61 WT lesions from 56 patients were eligible for inclusion, of which 30 lesions from 30 patients were included for direct correlation of IVIM-derived parameters and histopathology on slice-level (Figure 1). Twenty-six lesions were excluded due to a non-WT pathological diagnosis, partial nephrectomies not suitable for direct correlation using a cutting guide, and missing imaging or pathology data. The median age of the patients analyzed in this study was 44 months (range 4-105) (Table 1). All included lesions were from patients who received preoperative chemotherapy following the UMBRELLA protocol. In total, 94 slices obtained from these 30 lesions were eligible for analyses. A total of 27 slices (29%) were annotated as stromal, 3 (3%) as blastemal, and 5 (5%) as epithelial. In addition, 27 slices (29%) displayed a mixed pattern, 30 (32%) were categorized as regressive-type, and 2 slices (2%) were found to be entirely necrotic.

*Table 1: Patient characteristics.*

**Number of patients**

| **Number of patients** | 30 |
|---|---|
| **Sex** | |
| Male | 16 |
| Female | 14 |
| **Age (months)** | |
| Mean | 44.8 |
| Range | 1-105 |
| **Number of lesions** | 30 |
| **Number of slices** | 94 |
| **Pathological subtype (lesions \| per slice)** | |
| Stromal | 7 \| 27 |
| Epithelial | 1 \| 5 |
| Blastemal | 0 \| 3 |
| Mixed | 15 \| 27 |
| Regressive | 7 \| 30 |
| Completely necrotic | 0 \| 2 |

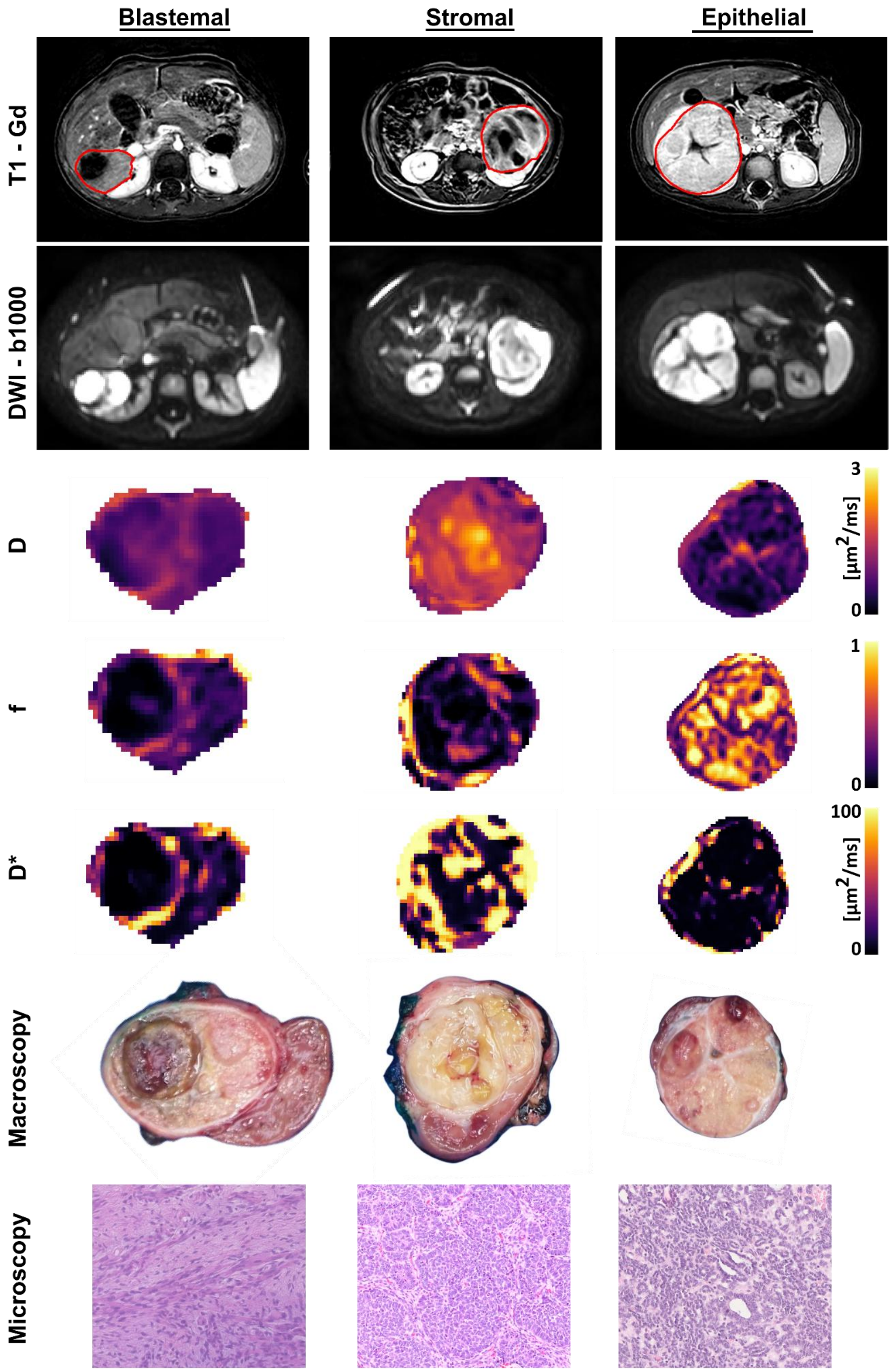

*Figure 2: Overview of MRI data and IVIM fitting results for three representative slices from the three different WT types.* **.** *The top row shows T1-weighted with gadolinium contrast (T1 – Gd) and DWI (b = 1000 s/mm²) MRI images with the tumor segmentation outlined in red. The middle row presents the IVIM parameter maps corresponding to the tumor region. The macroscopy row shows the corresponding macroscopic tissue slices used for histological validation. Bottom row shows corresponding histology on microscopy.*

### IVIM fitting

Results on IVIM fitting performance performed on simulated data can be found in the supplementary information (section S1). The largest performance increase for the INR fitting method has been shown for D* (r = 0.81 and RMSE = 13.74 versus r = 0.62 and RMSE = 24.38 for a segmented fitting approach). An overview of the input data, IVIM parameter results and corresponding macroscopy images are presented in Figure 2. Three macroscopic slices and microscopic sections from subtype stromal, epithelial and blastemal were selected as representative.

### Mixed model analysis

Median values for the three IVIM-derived parameters per pathological diagnosis are visualized as boxplots (Figure 3). Stromal tumors showed significantly higher D*-values than epithelial ($p = 0.024$), regressive tumors ($p = 0.031$) and mixed WT tumors ($p = 0.043$). No significant differences were found for median D or median f between any subtype. Median ADC showed no significant difference between stromal tumors and any other subtype in the mixed model analysis.

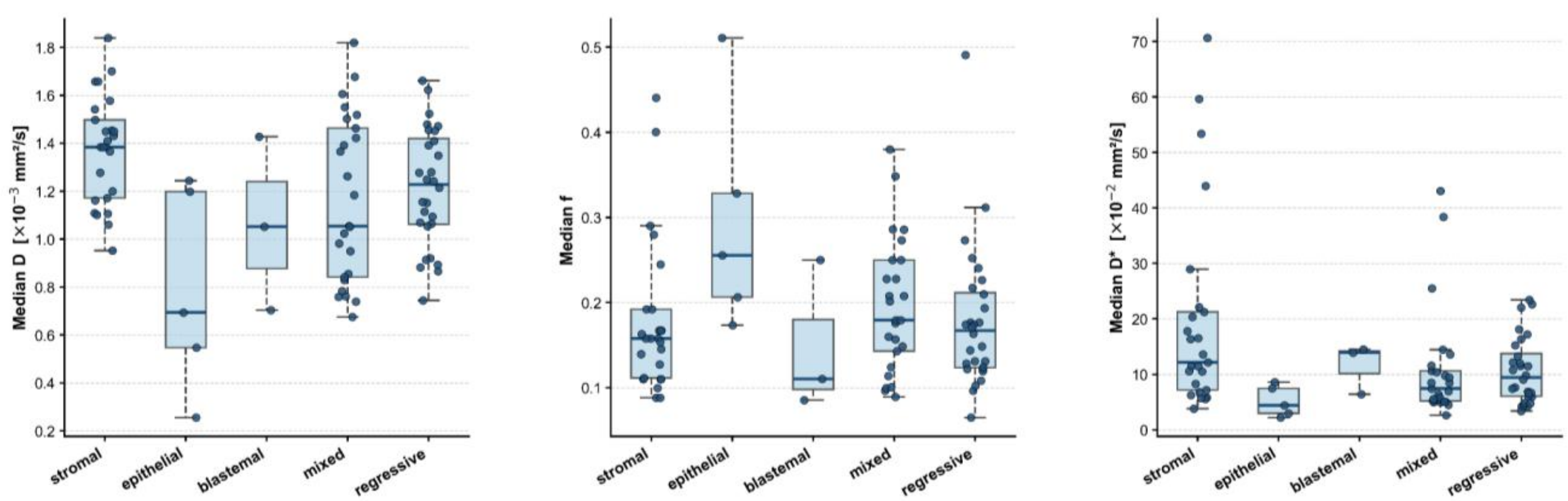

*Figure 3: Box-and-whisker plots show the median D (**A**), median f (**B**) and median D* (**C**) values per slice per pathological diagnosis after pre-operative chemotherapy for WT lesions on a macroscopic slice level (n=94 slices). Showing the difference in value distribution between the different pathological diagnoses.*

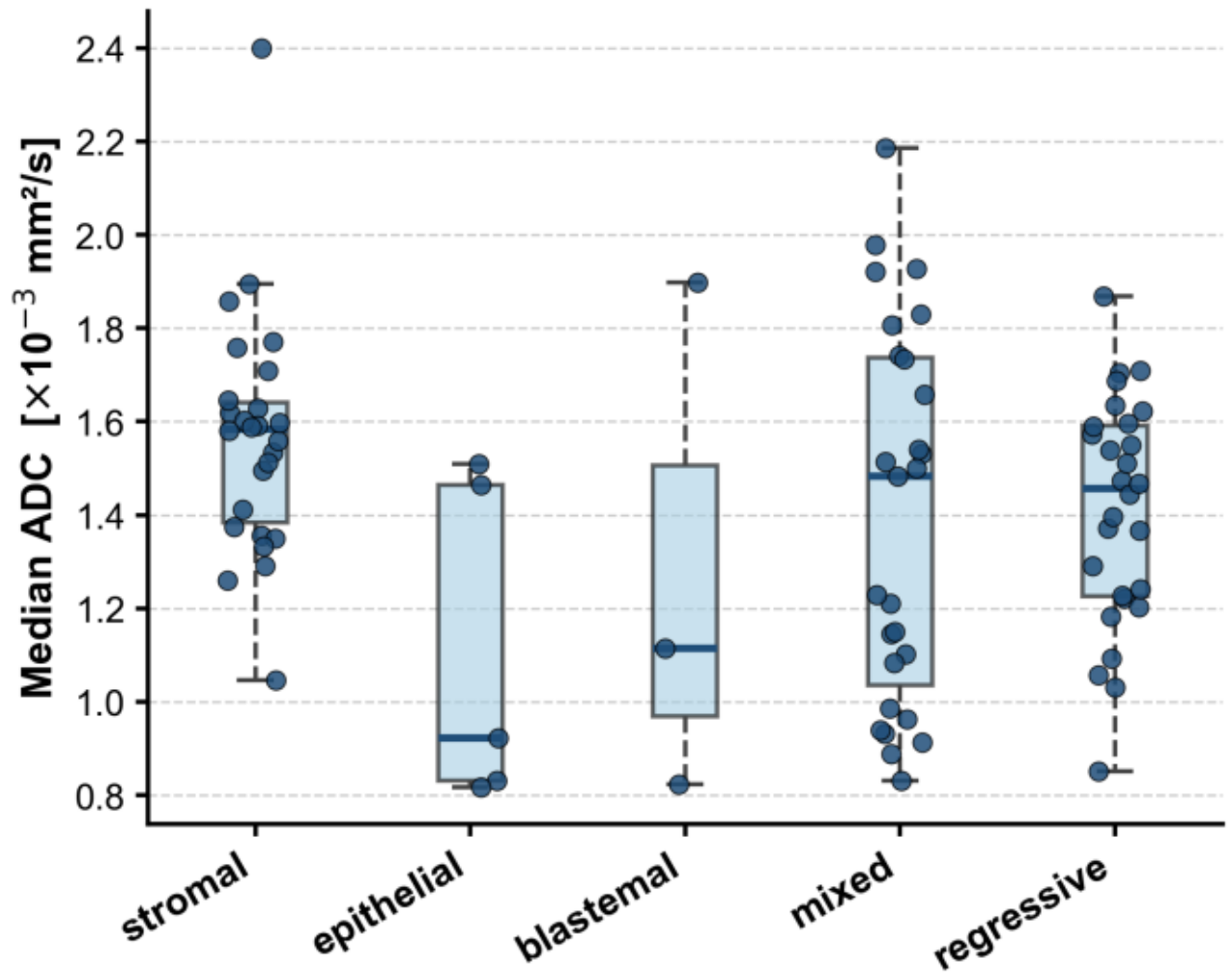

*Figure 4: Box-and-whisker plot shows the median ADC value per slice per pathological diagnosis after pre-operative chemotherapy for WT lesions on a macroscopic slice level (n=94 slices). Showing the difference in value distribution between the different pathological diagnoses.*

## Discussion

The aim of this study was to evaluate the discriminative performance of IVIM-derived parameters in differentiating WT subtypes on preoperative MRI. For this reason, we focused on the direct correlation of IVIM-derived parameters and histopathology in WT with a patient-specific cutting guide.

Mixed model analysis showed a significant difference between stromal and epithelial, mixed and regressive specimens, in median D*-values, suggesting that D* may hold discriminative value for identifying the blastemal components. This supports earlier studies which have found significant discriminating performance of D* [31], [32]. The improved accuracy of the INR fitting method for D* shows the importance of advanced fitting methods for IVIM. However, the small subset of slices for epithelial and blastemal subtypes prohibits drawing definitive conclusions. Additional slices are needed to further support these correlations. Other IVIM parameters as well as ADC values did not show significant findings.

The discrepancy with previous work [16], which reported a direct correlation between ADC and histopathology in WT, may reflect differences in histological processing. Beek et al. [16] correlated smaller histological samples with corresponding MRI regions of interest, whereas we correlated whole histological and MRI slices. Although the latter approach is more specific, it needs histological information about specific regions in the MRI, something which is only present after surgery such that WT subtype diagnosis using diffusion MRI would be subjective to a region of choice.

In our study, the patient distribution over WT subtypes is biased and strongly influenced by preceding chemotherapy, yielding less blastemal type WT and more mixed and regressive type WT [33]. The chemotherapy resistance of the epithelial and blastemal tissues present in this dataset could also make them not representative to chemotherapy sensitive versions of these tissue types. To be able to accurately predict subtypes based on imaging biomarkers, a correlation of diagnostic imaging and untreated tumors may be considered for example within the COG treatment protocol, biopsy tissue or lesions from children younger than six months.

This study has several limitations. First, DWI sequencing is susceptible to artefacts caused by long echo times and strong gradient pulses in combination with respiratory motion. Recent work has shown that this has a significant impact on DWI parameter estimation and classification in cancer [34]. Future work may focus on a dedicated postprocessing pipeline for pediatric renal DWI to correct for motion, eddy currents and susceptibility artefacts to improve fitting performance, which could subsequently improve correlation of WT subtype and IVIM parameters.
Secondly, although the cutting guide created a better correlation between MR imaging and histopathology, deformation of the tumor after resection could still introduce errors in the comparison to imaging. Isotropic MRI-DWI resolutions might allow for deformable 2D-3D registration in the future. Finally, our assessment on slice level of these different components without necrosis may not represent the overall pathological diagnosis of the patient on lesion level, used for treatment stratification.

## Conclusion

Direct correlation of histopathology and IVIM MRI resulted in a good differentiation between stromal subtype and epithelial, mixed and regressive subtype on macroscopic slice level using D*. Future research should therefore focus on the addition of data and equalization of the representation of different tissue types (mainly for blastemal and epithelial subtype), possibly by including chemo-naive tumors.

## Author contribution

Conception: GCA, MF, JH, ASL, CMWT;
Dataset generation: JNB, JH, RRK, MMEH, ASL;
Analysis and interpretation of data: GCA, JH, MF, JH, ASL, CMWT;
Drafting manuscript: GCA, MF, JH;
Revising manuscript: GCA, MF, JH, ASL, CMWT, AFWS, RRK, JNB, FFJS;
Funding: ASL, RRK, CMWT, AFWS;
Supervision: FFJS, EV, ASL, CMWT, AFWS, RRK.
Every author has approved the submitted version and is personally accountable for their own contributions.

## Acknowledgements

We thank the members of the surgical and pathological team involved in this study, as well as the other physicians involved. We thank the patients and parents for participating in the SIOP-RTSG 2016 UMBRELLA study. We thank Tom Hendriks for help with developing the INR fitting code and methods.

## Funding

This study has received funding from a grant (grant number 341) from the Stichting Kinderen Kankervrij (KiKa). GCA was supported by Eurostars 3 Project "FEM-SCAN: Fast and Efficient MRI Scanning for breast cancer detection"; Project id 3325. CMWT was supported by the Dutch Research Council (NWO) under a VIDI grant with file number 21299.

## Study or cohort overlap

Part of this cohort is previously published in European Radiology (DOI: 10.1007/s00330-024-10959-2) where ADC values were correlated with histopathology.